\documentclass[twocolumn,a4paper,showpacs,preprintnumbers,graphicx,amsmath,amssymb,reprint,floatfix]{revtex4-1}
\pdfoutput=1
\usepackage{graphicx}
\usepackage{dcolumn}
\usepackage{bm}
\usepackage{times}
\usepackage{multirow}
\usepackage{longtable}
\usepackage{natbib,hyperref}
\usepackage{color}

\begin{document}


\title{Exploring the subsurface atomic structure of the epitaxially grown phase change material Ge$_2$Sb$_2$Te$_5$}
\author{J. Kellner$^1$}
\author{G. Bihlmayer$^2$}
\author{V. L. Deringer$^3$}
\thanks{Present address: Engineering Laboratory, University of Cambridge, Cambridge CB2 1PZ, United Kingdom}
\author{M. Liebmann$^1$}
\author{C. Pauly$^1$}
\author{A. Giussani$^4$}
\thanks{Present address: Nano Power Research Laboratories, Rochester Institute Of Technology, Rochester NY 14623, USA}
\author{J. E. Boschker$^4$}
\author{R. Calarco$^4$}
\author{R. Dronskowski$^3$}
\author{M. Morgenstern$^1$}
\email[] {mmorgens@physik.rwth-aachen.de}
\affiliation{ $^1$II. Physikalisches Institut B and JARA-FIT, RWTH Aachen University, D-52074 Aachen, Germany}
\affiliation{ $^2$Peter Gr{\"u}nberg Institute and Institute for Advanced Simulation, Forschungszentrum J{\"u}lich and JARA, D-52428 J{\"u}lich, Germany}
\affiliation{ $^3$Institute of Inorganic Chemistry, RWTH Aachen University, D-52074 Aachen, Germany}
\affiliation{ $^4$Paul Drude Institut f{\"u}r Festk{\"o}rpelektronik Berlin, D-10117 Berlin, Germany}

\date{\today}

\begin{abstract}
Scanning tunneling microscopy (STM) and spectroscopy (STS) in combination with density functional theory (DFT) calculations are employed to study the surface and subsurface properties of the metastable phase of the phase change material Ge$_{2}$Sb$_{2}$Te$_{5}$ as grown by molecular beam epitaxy. The (111) surface is covered by an intact Te layer, which nevertheless allows to detect the more disordered subsurface layer made of Ge and Sb atoms. Centrally, we find that the subsurface layer is significantly more ordered than expected for metastable  Ge$_{2}$Sb$_{2}$Te$_{5}$. Firstly, we show that vacancies are nearly absent within the subsurface layer. Secondly, the potential fluctuation, tracked by the spatial variation of the valence band onset, is significantly less than expected for a random distribution of atoms and vacancies in the subsurface layer. The strength of the fluctuation is compatible with the potential distribution of charged acceptors without being influenced by other types of defects. Thirdly, DFT calculations predict a partially tetrahedral Ge bonding within a disordered subsurface layer, exhibiting a clear fingerprint in the local density of states as a peak close to the conduction band onset. This peak is absent in the STS data implying the absence of tetrahedral Ge, which is likely due to the missing vacancies required for structural relaxation around the shorter tetrahedral Ge bonds. Finally, isolated defect configurations with a low density of $~10^{-4}$/nm$^2$  are identified by comparison of STM and DFT data, which corroborates the significantly improved order in the epitaxial films driven by the build-up of vacancy layers.
\end{abstract}
\maketitle

\section{Introduction}
Phase change alloys are commercially used for optical data storage (DVD-RW, Blu-ray Disc) and for electrically addressable phase-change random-access memories (PC-RAM) \cite{Wuttig2007,Wuttig2012}. They typically exploit the large contrast in electrical conductivity and optical reflectivity between the amorphous and the metastable rock-salt phase of materials based on Ge, Sb and Te such as Ge$_2$Sb$_2$Te$_5$ (GST-225) \cite{Ovshinsky1968,Yamada1987,Raoux2009,Lencer2011}. The switching between the amorphous and the metastable phase favorably can occur within nanoseconds \cite{Yamada1991,Loke2012} and at an energy cost down to 1\,fJ for a single cell \cite{Xiong2011}.

Further optimization, in particular of the PC-RAM application, requires a more detailed knowledge on the mechanisms leading to phase change and electrical contrast. This should eventually expose structure-property relationships to be employed for the optimization of materials and their combination into novel types of composites \cite{Lencer2008}. The ongoing miniaturization of cells, though, increases the influence of surfaces and interfaces, which might be different from their bulk counterparts in terms of the atomic arrangement \cite{Deringer2015,Konze2016}. Hence, the atomic structure and the resulting properties in these distinct areas must be explored in detail, too. Adequate tools are transmission electron microscopy (TEM) \cite{Lotnyk2016b}, atom probe tomography \cite{Kelly2007,CojocaruMirdin2017}, and scanning probe microscopy \cite{Weidenhof1999,Subramaniam2009}. Among these, scanning tunneling microscopy (STM) and spectroscopy (STS) provide the advantage that the electronic density of states (DOS) close to the Fermi level $E_{\rm F}$ is probed simultaneously, down to the atomic length scale with sub-meV resolution \cite{Pan1998,Wiebe2004}. This allows to directly determine the corresponding structure-property relationship between atomic arrangement and electronic structure in real space. The probed local DOS, portraying the electronic structure, is dominated by the surface layer, but is also influenced by deeper layers \cite{Wittneven1998}.

Here, we explore STM and STS on the prototype phase change material GST-225 \cite{Wuttig2007,Deringer2015}, grown epitaxially by molecular beam epitaxy (MBE) \cite{Katmis2011,Rodenbach2012,Bragaglia2014,Bragaglia2016,Cecchi2017}. This provides the first STM study on single crystalline phase change materials. High surface quality without contaminations is achieved either by transferring the samples by an ultra-high vacuum (UHV) shuttle to the STM \cite{Kellner2017} or via cleaning them by a dip in deionized water prior to the insertion into the STM chamber \cite{Zhang2010}.

It is known from X-ray diffraction (XRD) that our epitaxial films grow in the technologically relevant metastable phase \cite{Boschker2017}. Conventionally, this phase features the cubic rocksalt structure with alternating layers of Te and of a disordered mixture of Ge, Sb and vacancies (Vcs). Both layers are hexagonally close-packed \cite{Zhang2012,Zhang2015,Zhang2016}. The layers are stacked in ABC order along the [111] direction of the rock-salt structure \cite{Matsunaga2004,Matsunaga2006,Matsunaga2008}. A partial ordering in between the mixed Ge/Sb/Vc layers, however, leads to vacancy rich Ge/Sb/Vc layers and vacancy poor Ge/Sb/Vc layers within the epitaxial films. These layers are partly stacked in a regular sequence, as deduced from XRD and TEM \cite{Bragaglia2016}. This more ordered phase deviates from the rocksalt structure and has recently been dubbed the metastable vacancy-ordered phase \cite{Boschker2017}. Moreover, it is known that the epitaxial films are strongly p-type doped \cite{Pauly2013} with an unintentionally varying charge carrier density $p=(0.1-3)\cdot 10^{26}$~m$^{-3}$ \cite{Kellner2017, Bragaglia2016}. This doping is typically related to excess vacancies \cite{Wuttig2006,Edwards2006,Schaefer2017}. The epitaxial films also exhibit a significantly improved mobility $\mu$  with respect to polycrystalline films prepared by magnetron sputtering, i.e., an increase of $\mu$ by more than an order of magnitude \cite{Bragaglia2016,Volker2013,Kellner2017}. Since $\mu$ of the polycrystalline films barely depends on grain size \cite{Volker2013,Deringer2015}, this corroborates the improved order in the epitaxial films.

Here, we find, by comparison of STM and DFT data, that the epitaxial films are Te terminated. Hence, the more interesting disordered layer is the subsurface layer, which is more difficult to access by STM and STS.
Nevertheless, we observe spatial fluctuations in STM and STS data which can be traced back to the disordered subsurface layer.

Firstly, fluctuations of the onset of the valence band by about 20 meV on the length scale of about $1-2$ nm are found. Surprisingly, these potential fluctuations can be reproduced by a random distribution of screened acceptor potentials in the bulk of the film, which provide the charge carrier density $p=3\cdot 10^{26}$~m$^{-3}$ as deduced from Hall measurements of identically prepared samples. This points to an increased order of the subsurface region with respect to a completely disordered distribution of Ge, Sb and Vcs, in line with previous findings on the epitaxial films \cite{Bragaglia2016}.

Secondly, we show by DFT calculations that a totally disordered subsurface layer exhibits tetrahedral bonding of Ge, if the Ge is close to a Vc. Such tetrahedral bonding of Ge surrounded by Vcs has been conjectured some time ago also for the bulk of the GST-225 rock-salt structure by using electron microscopy and diffraction data in combination with DFT results \cite{Liu2011}. However, this interpretation has not been confirmed in subsequent works using TEM \cite{Lotnyk2016,Zhang2016,Sun2016}, XRD \cite{Kohara2006,Fons2012}, extended X-ray absorption fine structure (EXAFS) \cite{Kolobov2004,Krbal2012} and more refined DFT calculations, partially combined with molecular dynamics (MD) simulations \cite{Raty2012,Kalikka2014}. Interestingly, a nuclear magnetic resonance (NMR) study provided some hints for the presence of tetrahedral Ge within a so-called nanocrystalline GST-225 phase, but not in its microcrystalline counterpart. This would imply an increased tendency for tetrahedral bonding close to the surface of GST-225 \cite{Sen2012}. Our DFT data reveal that if tetrahedral bonding were to occur within the subsurface layer, this would lead to a clear fingerprint in the electronic structure, viz. a peak at the conduction band onset in the local density of states (LDOS). This peak has a strong $s$-type character and, hence, persists as a peak within the LDOS up to 7~\AA\hspace{1mm} above the surface, implying that it should be observable by STS. However, within the STS data, we do not find such a peak and, thus, rule out a significant presence of tetrahedral Ge in the subsurface layer of the epitaxial films.  We attribute this finding to the reduced presence of vacancies in the subsurface layer \cite{Bragaglia2016}, which are required for relaxation around the shorter tetrahedral Ge bond.

Finally, we compare the LDOS from DFT data of particular defect configurations with STM data. This reveals that isolated defects are observable within our structure, probably even defects being located several layers below the surface. This would not be possible for a completely disordered subsurface layer due to the overlap of multiple different LDOS fingerprints. Hence, again, we conclude that the subsurface layer is partially ordered providing only a few defects. We refrain from an exact assignment of the found patterns in STM images to atomic defect structures due to the large number of possible configurations in a disordered hexagonal layer consisting of three components.


\section{Sample Preparation and STM/STS Experiment}

Thin GST-225 films (thickness: 20 nm) were grown via MBE on a carefully cleaned Si(111) substrate at temperature $T=250$\,$^{\circ}$C \cite{Katmis2011,Rodenbach2012,Pauly2013}. The substrates are primarily prepared to reveal the Si(111)-$7\times 7$ reconstruction or the Si(111)$\sqrt{3}\times \sqrt{3}$-Sb reconstruction using slightly different protocols for substrate preparation \cite{Boschker2014}. Afterwards, GST-225 is deposited using distinct sources for each element.
XRD reveals that the GST-225 films grow epitaxially  exhibiting the single crystalline, meta-stable phase with [111] surface. Twin domains are deduced from XRD, i.e., adjacent areas of ABC and CBA stacking of the hexagonal layers \cite{Pauly2013}. The samples on the Si(111)-$7\times 7$ reconstruction additionally feature rotational domains.
An XRD peak indicating the formation of vacancy layers is observed attributing the epitaxial films to the metastable vacancy-ordered  phase \cite{Bragaglia2016,Boschker2017}.

STM and STS measurements are performed either using a home-built STM operating in UHV at $T=9$ K \cite{Liebmann2017} or a home-built room temperature UHV-STM setup \cite{Geringer2009}.  Samples probed at $T= 9$\,K have been prepared on Si(111)$\sqrt{3}\times \sqrt{3}$-Sb and are afterwards transferred by a UHV shuttle between the MBE and the STM system  at an average pressure $p = 5\times10^{-10}$\,mbar. Samples probed at $T=300$\,K have been grown on the Si(111)-$7\times 7$ reconstruction and are transported under ambient conditions to the STM chamber, but are dipped into de-ionized water for one minute directly before the insertion into the UHV chamber ($2-3$ min before pumping the load-lock). This procedure was followed by annealing at $200^{\circ}$C in UHV for half an hour \cite{Zhang2010}. Both methods reveal GST-225 surfaces free of oxides and other surface contaminations as visible in the STM data and cross-checked by x-ray photoelectron spectroscopy (XPS) and Auger electron spectroscopy. We did not observe any differences between the two preparation methods also regarding the topographic STM images. Since surface oxidation starts already at $\sim 10^4$\,L of O$_2$ \cite{Yashina2008}, transfer in UHV  or removal of the oxide is mandatory. XPS is also used to cross-check the stoichiometry of the GST-225 samples.

Topographic STM measurements are performed in constant-current mode applying the voltage $V$ to the sample. As STM tips, we use ex-situ etched W wires, which are additionally prepared in-situ by voltage pulses. To improve the image quality, we employed a Gaussian smoothing of the recorded images with a maximum lateral full width at half maximum of $0.6$\,\AA. We verified that the removed noise does not contain atomic scale features and that the vertical amplitude of the removed noise does not exceed 10 pm.
In order to directly compare consecutively recorded STM images at different $V$, we applied a drift compensation using point defects or kinks in step edges as track features.

To compare the constant current maps $z(x,y)$ from experimental STM images with calculated LDOS maps at constant distance $z$ from the surface,  we partly convert $z(x,y)$ into current  $I_{\rm STM}(x,y)$ as expected in constant-height mode using \cite{Tersoff1983,Tersoff1985}
\begin{equation} \label{eq:2}
I_{\rm STM}(x,y)=I_0 e^{-2\kappa z(x,y)}.
\end{equation}
The decay constant $\kappa$ is determined from $I(z)$ measurements ($I$:~measured current). We find $\kappa=9.4$\,nm$^{-1}$ for the measurements at $T=9$\,K and $\kappa =12$\,nm$^{-1}$ for the measurements at $T=300$\,K. The prefactor $I_0$ is taken as a spatially constant scaling factor.

We also estimate the absolute value of the tip-sample distance $z$ by $I(z)$ curves using the conductance quantum $G_{0}$ = $2e^{2}/h$ ($h$: Planck constant), which is assumed to be the conductivity at the tip-sample contact point $z=0$\,\AA. This leads to \cite{Kroeger2007,Mashoff2010}
\begin{equation} \label{eq:1}
z=-\frac{1}{2\kappa}\mathrm{ln}\left(\frac{1}{G_{0}}\frac{I}{V}\right),
\end{equation}

Spectroscopic $dI/dV(V)$ curves, which are proportional to the LDOS$(E)$ at energy $E$ \cite{Tersoff1983,Tersoff1985,Morgenstern2003}, are recorded with open-feedback loop after stabilizing the tip-sample distance at current $I_{\rm stab}$ and voltage $V_{\rm stab}$. We use lock-in technique with a sinusoidal modulation voltage of amplitude $V_{\mathrm{mod}}=5$\,meV, resulting in an energy resolution $E_{\mathrm{res}}=\sqrt{(3.3\cdot k_{\mathrm{B}}T)^{2}+(1.8\cdot eV_{\mathrm{mod}})^{2}}\approx 10$\,meV ($k_{\mathrm{B}}$: Boltzmann constant, $e$: elementary charge, $T=9$\,K) \cite{Morgenstern2003}.

The $dI/dV$ spectra are subsequently normalized twice. Firstly, we remove the influence of the remaining low-frequency mechanical vibrations \cite{Freitag2016}. In constant-current mode, these vibrations are compensated by the feedback-loop. Opening the feedback-loop, hence, stabilizes the tip-sample distance at an uncontrolled phase of the oscillation. The ongoing oscillation then leads to a different average distance between tip and sample than intended. Consequently, the time averaged current $\overline{I}(V_{\rm stab})$, which is recorded after opening the feedback loop, differs from $I_{\rm stab}$. This effect is compensated by dividing the recorded $dI/dV(V)$ by $\overline{I}(V_{\rm stab})$ \cite{Freitag2016}. Secondly, the tip-sample distance at $(I_{\rm stab}, V_{\rm Stab})$ depends on the lateral position $(x,y)$ due to the spatially varying LDOS$(E,x,y)$ \cite{Morgenstern2003}. Since we are interested in the LDOS probed at a spatially constant tip-sample distance $z$, we compensate the varying $z$ by independently recording $I(z)$ curves and the local $z(x,y)$, i.e., a constant-current image. Dividing $dI/dV(V,x,y)$ by $I(z(x,y))$ rescales different $dI/dV$ curves, as if they have been measured at a constant $z(x,y)$   \cite{Wittneven1998}. For the sake of simplicity, we call the doubly renormalized curves $dI/dV_{\rm scaled}(V)$.


\section{Density Functional Theory}

DFT calculations without spin-orbit coupling (SOC) are performed using the PBE-GGA parametrization \cite{Perdew1996} and the VASP implementation \cite{Kresse1996,Kresse1996b,Kresse1999} of the projector augmented plane wave method \cite{Bloechl1994} as described in more detail elsewhere \cite{Kellner2017}. Ge 4s 4p, Sb 5s 5p, and Te 5s 5p states were expanded into plane waves up to an energy cutoff of $250$ eV, and $k$ space was sampled using a $\Gamma$-centered $1 \times 1 \times 3$ grid (the $\Gamma$ point only) for bulk (surface) cells, respectively. Firstly, we modeled the bulk of meta-stable rock-salt GST-225, where the Te atoms are assumed to form a hexagonal, defect free layer alternating with a hexagonal layer of Ge, Sb and vacancies in random order (Fig. \ref{Fig3}a). The hexagonal unit cell is repeated $5 \times 5 \times 1$ times in the $a \times b \times c$ directions. The so-obtained supercell hosts three layers for Ge, Sb and vacancies, each with $25$ possible atomic sites, where $10$ Ge, $10$ Sb and 5 vacancies are distributed randomly, while the other sublattice is filled with Te atoms.  Finally, inversion symmetry with regard to the cell center is imposed. We set up three different bulk configurations each with a different randomized distribution of Ge, Sb, and Vc. Furthermore, copies of these cells were made with exchanged Ge and Sb positions. The six bulk models were structurally optimized while the cubic cell shape was enforced. The bulk simulations are cross-checked by comparison with experimental data. After relaxation, the bulk unit cells have densities of $\rho_{\mathrm{model}} = 0.0311-0.0313$\,atoms/\AA$^{3}$ (experiment: $\rho = 0.033$ atoms \AA$^{-3}${} \cite{Steimer2008}) The rock salt lattice parameters are $a_{\mathrm{model}}=6.127-6.139$\,\AA{} (experiment: $a = 6.0245(1)$\,\AA{} \cite{Yamada2000}). The small difference reflects the typical small underbinding within GGA based calculations \cite{Silva2008}.

The (111) surface is modeled by cutting symmetric slabs of 15 and 17 layer thickness for Te- and GeSb-terminated surfaces, respectively, from the relaxed hexagonal bulk cells followed by additional relaxation of all atoms. Te-terminated surfaces are mainly considered, as they are more stable by about 50 meV/\AA$^2$, as numerically validated for more ordered unit cells previously \cite{Deringer2013}.

For the calculations with SOC, we used the relaxed structure from the calculations without SOC and employed the full-potential linearized
augmented plane wave (FLAPW) method \cite{Wimmer1981} implemented in the FLEUR code (www.flapw.de),
using the same exchange-correlation functionals as
described above. Comparing the density of states (DOS) with and without
SOC, we find only a small reduction of the band gap due to the level splittings
by SOC. Notably, no topological surface states  were found in the gap, in contrast to more ordered configurations of GST-225 \cite{Kellner2017,Pauly2013,Silkin2013}.

Besides, we also employed DFT calculations with single defects using the same code as described above without SOC. As a starting configuration we used a $5 \times 5$ times repetition of an ideal, defect-free GST-225 (0001) surface based on the Kooi-de Hosson stacking \cite{Kooi2002} and implemented the defects by removal or exchange of atoms prior to relaxation. The local density approximation (LDA) was employed as our previous work showed that the pure GGA does not yield reliable surface energies for well-ordered GST-225(0001) surfaces \cite{Deringer2013}.

We calculated the LDOS in vacuum to simulate STM and STS measurements directly, using the FLAPW method in thin-film
geometry \cite{Krakauer1979,Ferriani2010}. STM images are described according to the Tersoff/Hamann model \cite{Tersoff1983, Tersoff1985}, i.e. we integrate the LDOS$(E)$ from the energy $E=E_{\rm F}$ to  $E=E_{\rm F}+eV$ at a distance $z_{\rm DFT}$ from the surface. The surface position is therefore defined as the average position of the surface atoms. We call this measure $I_{\rm DFT}:=\int_{E_{\rm F}}^{E_{\rm F}+eV} {\rm LDOS}(E) dE$, since it mimics the current map expected in constant height STM images. The resulting $I_{\rm DFT}(x,y)$ is displayed using a logarithmic scale in order to ease the comparison with the experimental STM data, which are measured in constant current mode (see eq.~\ref{eq:2}). With the help of the experimental $I(z)$ curves, spatial fluctuations of $I_{\rm DFT}(x,y)$ at constant $z_{\rm DFT}$ can be converted into calculated, spatial $z_{\rm DFT}(x,y)$ fluctuations as expected in constant-current mode STM images.

\section{Te termination and subsurface vacancies}

\begin{figure}
\includegraphics[width=\linewidth]{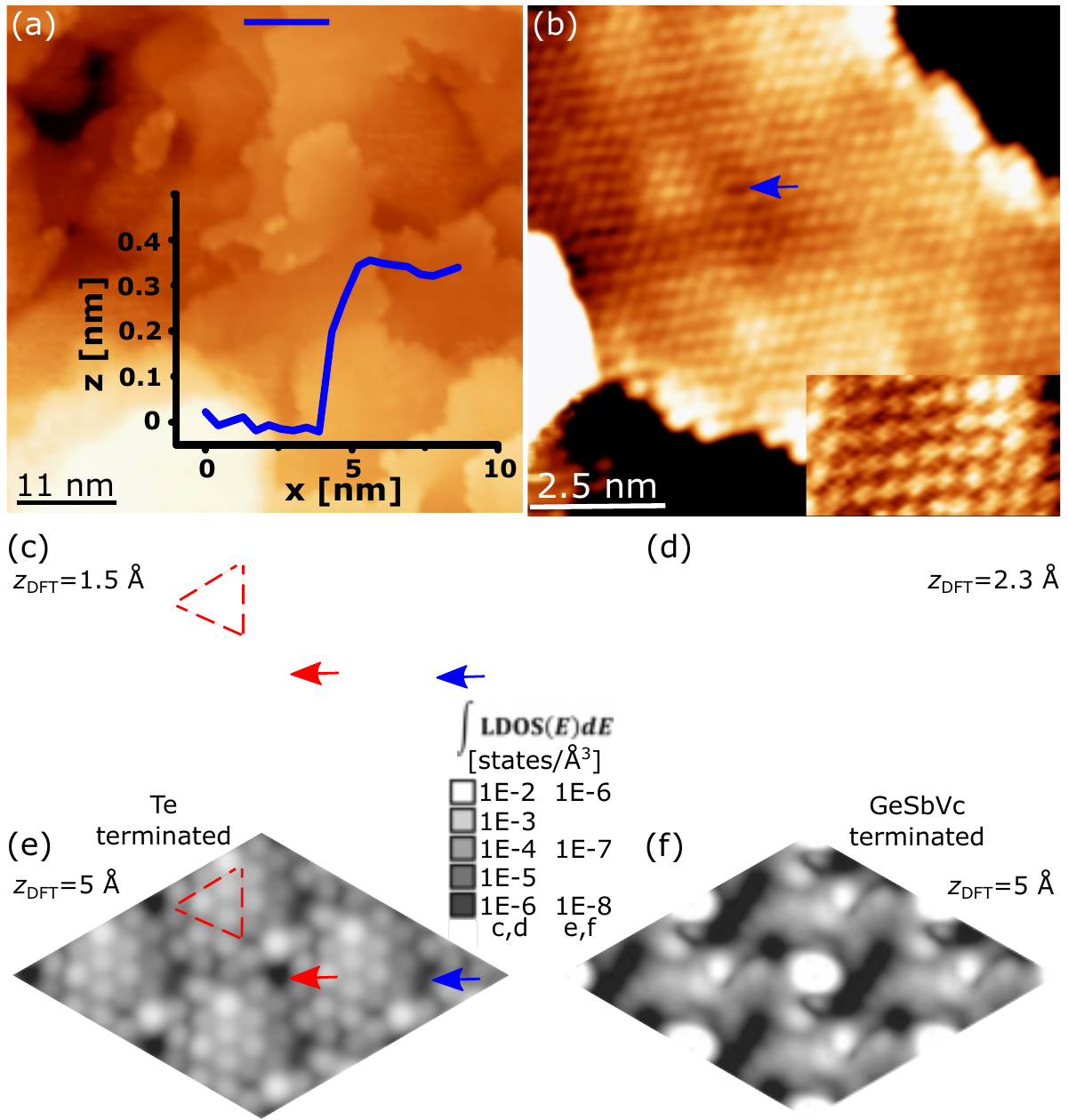}
\caption{(a) Large scale STM image of GST-225, $V=-0.3$\,V, $I = 100$\,pA, $T=300$\,K; blue line marks the profile line shown in the inset. (b) Atomically resolved STM image with inset at larger magnification, $V=-0.5$ V, $I=100$ pA, $T=300$\,K, blue arrow marks a dark spot in the apparent atomic lattice. (c) DFT simulated STM images of GST-225 with Te termination including SOC, $z_{\rm DFT}=1.5$\,\AA, $V=-0.35$\,V, red (blue) arrow marks an exemplary darkest (less dark) spot in the apparent atomic lattice (see text); red triangle marks vacancy surrounded by three octahedrally bonded Te atoms. (d) DFT simulated STM image of GST-225 with Ge/Sb/Vc termination including SOC, $z_{\rm DFT}=2.3$\,\AA, $V=-0.35$\,V. (e) Same as (c), but at $z_{\rm DFT}=5.0$\,\AA. (f) Same as (d), but at $z_{\rm DFT}=5.0$\,\AA. The logarithmic color scale for (c)$-$(f) shows the integrated LDOS from $E_{\rm F}$ to $-0.35$\,eV; a.u.: atomic units. \label{Fig1}}
 \end{figure}

Figure~\ref{Fig1}a shows a large scale STM image of the epitaxial GST-225(111) surface. Atomically flat terraces with widths in the $10$\,nm range and heights of $3.4 \pm 0.1$\,\AA{} are found \cite{Pauly2013}. This corroborates that the films are in the metastable phase, which exhibits a distance between adjacent Te layers of $3.47$\,\AA{} \cite{Nonaka2000}. Atomically resolved STM images (Fig.~\ref{Fig1}b) show a regular hexagonal pattern with corrugations in the sub-\AA{} regime of a correlation length of $2-3$ atomic distances. Similar images are found everywhere on the sample after both types of transfer (see also Fig.~\ref{Fig2}a).

Within images of $10\times10$\,nm$^2$, the corrugation exhibits a Gaussian distribution with $\sigma$-width $\sigma=0.25 \pm 0.05$\,\AA. For comparison, Fig.~\ref{Fig1}c$-$f displays  simulated STM images for a GST-225 film with randomly disordered Ge/Sb/Vc layers employing DFT without SOC. They are displayed at two different $z_{\rm DFT}$. For each case, they are shown for the energetically favorable Te termination of the film (Fig.~\ref{Fig1}c, e) and without the Te top layer, which is artificially removed after relaxation (Fig.~\ref{Fig1}d, f). It is obvious that the regular hexagonal atomic arrangement observed in the experiment is only compatible with the Te termination. The simulated STM images at larger, more realistic $z_{\rm DFT}$ are slightly more blurred than in the experiment. This is probably caused by the presence of tip orbitals with larger angular momentum, typically present in experiments using a W tip, and improving the atomic resolution \cite{Chen1990}.

The corrugation of the simulated, Te-terminated STM images is converted into a $z_{\rm DFT}(x,y)$ corrugation, which exhibits a Gaussian distribution with $\sigma=0.5$\,\AA, rather independent on the spatially averaged $\rm z_{\rm DFT}$. This reasonably fits with the experiment ($\sigma=0.25$\,\AA), albeit being slightly larger. The latter is caused by several dark spots in the calculated atomic lattice (arrows in Fig.~\ref{Fig1}c, e), which are barely found in the experiment (arrow in Fig.~\ref{Fig1}b). The darkest spots in the simulated images exhibit a reduction of $I_{\rm DFT}$ by a factor of about 15 with respect to the average value, which implies an apparent depth in constant-current images of $\sim 1.4$\,\AA{} \cite{Morgenstern2003}.
The dark spots are centered at Te positions of the surface layer, which are adjacent to Vc positions in the underlying Ge/Sb/Vc layer.
The deepest depressions, exemplary marked by red arrows in Fig.~\ref{Fig1}c, e, are surrounded by two subsurface Vcs. The corresponding atoms are relaxed downwards by $1.2$\,\AA.
Less deep depressions (blue arrows in Fig.~\ref{Fig1}c,e) are surrounded by a single subsurface Vc, but exhibit tetrahedral bonding to a neighboring subsurface Ge atom, in addition. They are moved downwards by $1.3$\,\AA, but appear less deep due to the modified LDOS by the different bonding (see below).
We conclude that dark Te atoms indicate the presence of certain types of subsurface vacancies. We find that $\sim 2/3$ of the subsurface Vcs result in a depression in the simulated STM images. The remaining $1/3$ of subsurface Vcs (exemplary surrounded by a red triangle in Fig.~\ref{Fig1}c,e) are surrounded by octahedrally bonded surface Te atoms. These Te atoms are barely relaxed in vertical direction and even appear slightly brighter than the surrounding Te atoms.

Since the atomic resolution in the experiment appears sharper than in the DFT results (Fig.~\ref{Fig1}b,e), we rule out that our tip is too blunt to observe the atomic scale depressions.
Hence, we conjecture that the subsurface layer contains significantly less Vcs than expected for a totally mixed Ge/Sb/Vc layer. This is in line with the observed XRD peak of these films, indicating the formation of separate Vc layers, such that the remaining Ge/Sb/Vc layers become vacancy poor \cite{Bragaglia2016}.

\section{Potential fluctuations at the surface}

The fact that the surface is Te terminated could be regarded as detrimental for STM investigations, since the more interesting Ge/Sb/Vc layer is subsurface.
However, Fig.~\ref{Fig1}f reveals that a termination by a Ge/Sb/Vc layer would result in rather irregular STM images, probably difficult to interpret.
Moreover, fingerprints of the disordered subsurface layer can be obtained by STM and STS still.

\begin{figure}
\includegraphics[width=1\linewidth]{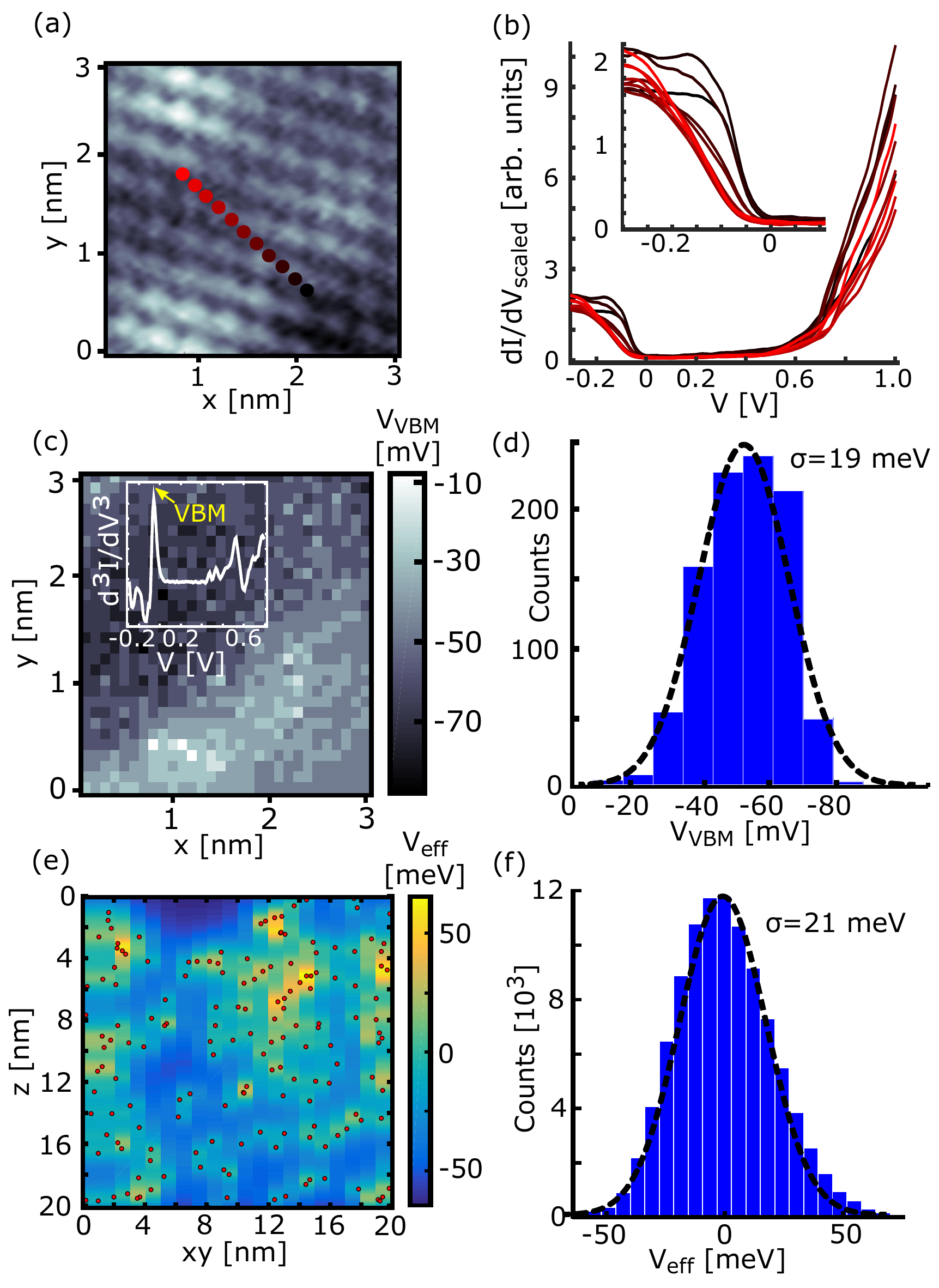}
\caption{(a) Atomically resolved STM image of GST-225 after UHV transfer, $V=-300$~mV, $I=50$~pA, $T=9$~K, colored points mark positions of spectra shown in (b). (b) Scaled $dI/dV(V)$ spectra recorded at the positions marked in (a) with zoom into the region of the valence band maximum (VBM) as inset, $V_{\rm stab}=-300$~mV, $I_{\rm stab}=50$~pA, $T=9$~K. (c) Map of the voltage at VBM ($V_\mathrm{VBM}$) determined as the local maximum close to the VB onset in $d^3I/dV^3(V)$ curves (inset), same area as in (a). (d) Histogram of $V_\mathrm{VBM}$ (blue bars) with dashed, Gaussian fit curve of $\sigma$-width as marked. (e) Vertical cut through the simulated electrostatic potential $V_{\mathrm{eff}}(x,y,z)$ for randomly distributed bulk acceptors (red dots) at density $N_{\rm A}=3\cdot 10^{26}$/m$^3$. (f) Histogram of the potential values $V_{\mathrm{eff}} (x,y)$ at the surface resulting from multiple simulations (blue bars). Gaussian fit curve with marked $\sigma$-width is added as a dashed line. \label{Fig2}}
 \end{figure}

Firstly, we investigate potential fluctuations, which are partly caused by charged defects (acceptors within GST-225) surrounded by a long-range screened Coulomb potential. A possibility to track the local potential
$V_{\rm eff}(x,y)$ within the surface layer is to determine the spatially varying valence band maximum (VBM) \cite{Salemink1991,Weidlich2011}. Figure~\ref{Fig2}a$-$b indeed show that the VB onset varies on the sub-nm length scale. The scaled $dI/dV$ curves also reveal that the VBM, exhibiting a step-like appearance, is more clearly defined in the LDOS than the conduction band minimum (CBM). A reasonable measure of the VBM is given by the inflection point of the LDOS, which is the local maximum of $dI^3/dV^3(V)$ (inset of Fig.~\ref{Fig2}(c)). It is displayed as a function of position in Fig.~\ref{Fig2}c.
The continuous spatial evolution implies that the maximum indeed tracks $V_{\rm eff}(x,y)$ of the surface layer. The corrugation observed in this small area reveals a Gaussian distribution with $\sigma \simeq 20$\,mV (Fig.~\ref{Fig2}d). The correlation length $\xi$ is determined by the $1/e$ decay length of the angularly averaged correlation function from Fig.~\ref{Fig2}c. It turns out to be $\xi = 1.8$\,nm.
The mean value $\overline{V}_{\rm VBM} \simeq -60$\,mV is in reasonable agreement with the previously determined VBM by angular resolved photoelectron spectroscopy (ARPES), being at $E-E_{\rm F}=-105\pm 10$\,meV  \cite{Kellner2017}. We note that ARPES is more precise concerning the absolute value of the VBM due to its wave vector resolution. Nevertheless, the STM data corroborate that the VBM is well below $E_{\rm F}$ for epitaxial GST-225 within the metastable vacancy-ordered phase \cite{Kellner2017}. In addition, we consistently observe a finite $dI/dV_{\rm scaled}$ within the band gap (Fig.~\ref{Fig2}b), i.e., in the energy region of $\simeq 0.5$\,eV above the VBM \cite{Lee2005,Pauly2013,Kellner2017}. We attribute this LDOS to the topological surface state found recently by 2-photon ARPES \cite{Kellner2017}.\\

For comparison, we simulate the electrostatic potential $V_{\rm eff}(x,y,z)$ within GST-255 using randomly distributed bulk acceptors employing a simulation, which is described in detail elsewhere \cite{Bindel2016}. As a lower bound for the randomly distributed bulk acceptor density $N_{\rm A}$, we use the measured p-type charge carrier density of identically prepared GST-225 films $N_{\rm A}= p = 3\times10^{26}$/m$^{3}$ \cite{Bragaglia2016,Kellner2017}. Each acceptor is described as a screened Coulomb potential
\begin{equation}
V_{\rm Coul}(\underline{r})=\frac{e}{4\pi \epsilon_{0}\epsilon_{\rm r}|\underline{r}|} \exp{(-|\underline{r}|/\lambda)}
\end{equation}
with the dielectric constants $ \epsilon_{0}$ of vacuum and $ \epsilon_{\rm r}$ of GST-225, the vector from the center of the acceptor $\underline{r}$, and the screening length $\lambda$, being
$\lambda^{-2}=4(\frac{3}{\pi})^{\frac{1}{3}}p^{\frac{1}{3}}/a_{\rm B}$ with the Bohr radius $a_{\rm B}=4\pi \hbar^{2} \epsilon_{0} \epsilon_{\rm r} /m_{\mathrm{eff}} e^{2}$.  The static dielectric constant of GST-225 is $\epsilon_{\rm r} = 33.3$ \cite{Shportko2008} and the effective mass is $m_{\mathrm{eff}}=0.35 \cdot m_{\mathrm{e}}$ ($m_{\mathrm{e}}$: bare electron mass)  \cite{Kellner2017}. This results in $\lambda = 1.4$\,nm.

We simulate a volume of $20\times140\times 140$ nm$^{3}$ containing about $10^5$ acceptors using a pixel grid of 1 nm$^3$. The sample thickness of $20$ nm is chosen as in the experiment. To avoid boundary effects, we evaluate the resulting $V_{\rm eff}(x,y)$ at the surface only in the central surface area of $20\times 20$\,nm$^2$. A cut through the resulting potential with marked nearby acceptors is shown in Fig.~\ref{Fig2}e. The histogram of $V_{\rm eff}(x,y)$ resulting from $\sim 100$ simulation runs exhibits a Gaussian distribution with  $\sigma = 21$ meV (Fig. \ref{Fig2} (f)), very close to the value found in the experiment ($\sigma \simeq 20$\,mV). The correlation length of the simulation is $\xi = 1.6$\,nm again very close to the experimental value ($\xi \simeq 1.8$\,nm).\\
Hence, albeit we do not have enough STS data to draw statistically firm conclusions, it is obvious that the bulk acceptors largely explain the observed potential fluctuations. In turn, potential fluctuations resulting from the additional disorder in the subsurface layer, being characteristic for sputtered, cubic GST-225 films \cite{Zhang2012,Zhang2015,Zhang2016}, barely matter. This is partly due to the more short-range character of the potential surrounding this type of uncharged disorder, but, more importantly, it reveals again the reduced disorder within the subsurface layer.

The latter is deduced straightforwardly from the comparison with the potential fluctuations of the simulated GST-225 film having completely randomized Ge/Sb/Vc layers. We again use the inflection point of the LDOS at $z_{\rm DFT} = 5$\,\AA{} to pinpoint the energy of the VBM. The spatial distribution of this energy turns out to exhibit a Gaussian with $\sigma \simeq 50$\,meV for the area displayed in Fig. \ref{Fig1}c$-$f. This is significantly larger than in the experiment ($\sigma \simeq 20$\,mV).

\section{Tetrahedral bonding}

\begin{figure}
\includegraphics[width=1\linewidth]{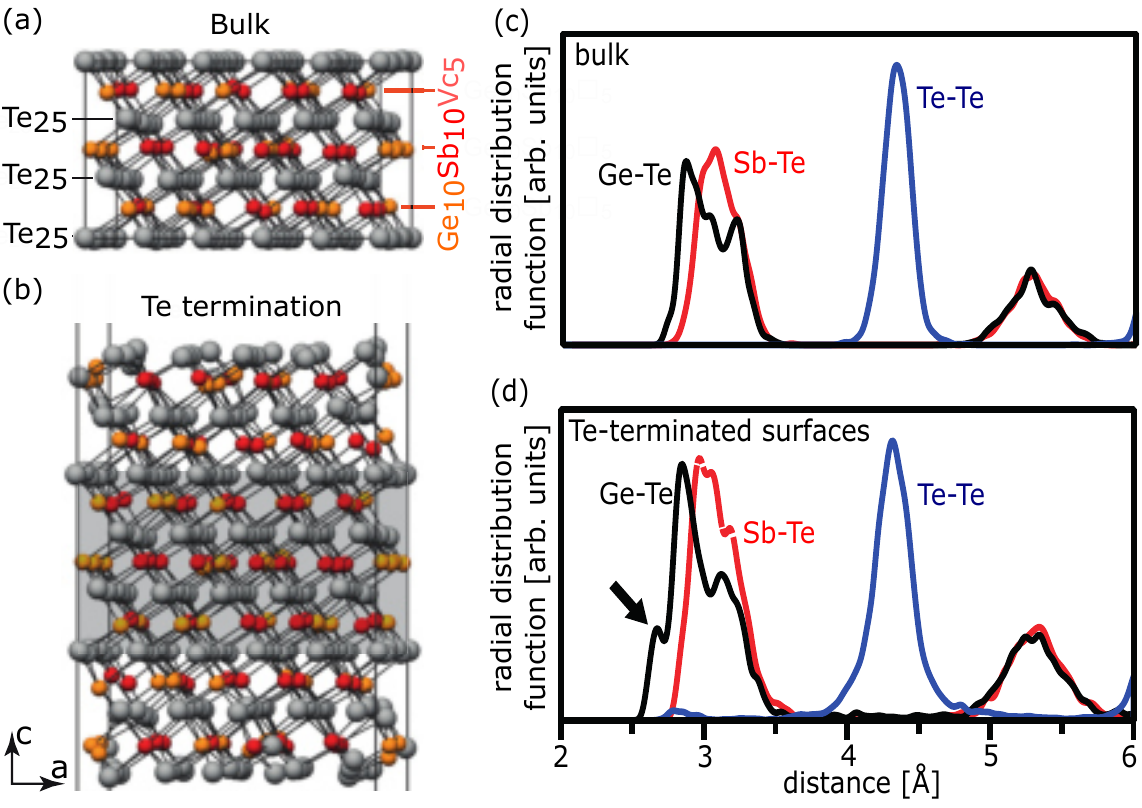}
\caption{(a) Relaxed atomic positions of metastable rock-salt phase of GST-225 of a bulk unit cell: Te atoms (grey), Sb atoms (red),
  Ge atoms (orange). The atomic composition of the layers is marked. (b) Relaxed atomic positions in a slab of metastable GST-225 terminated by Te surface layers (space group \textit{P}$\overline{1}$, same color code as in (a)). The central bulk-like area is shaded in grey. Note that all atomic positions have been relaxed. (c) Radial distribution functions of labeled interatomic distances deduced from several bulk-type calculations as displayed in (a). (d) same as (c), but for the models with Te terminated surfaces as displayed in (b); arrow marks Ge-Te distance of tetrahedral Ge bonding; a Gaussian smoothing has been applied to ease visualization. \label{Fig3}}
 \end{figure}

A strong relaxation of the surface Te atoms next to vacancies is apparent through the dark spots in Fig. \ref{Fig1}c and e. This might also change the bonding distance of Te atoms to the neighboring Ge and Sb atoms within the Ge/Sb/Vc layer. The resulting shorter bond length and bonding asymmetry points to a resulting tetrahedral bonding configuration \cite{Liu2011}. Note that tetrahedral Ge bonds appear to reveal the strongest bonds in amorphous GST-225 \cite{Lee2017}. Indeed, we find tetrahedral bonding configurations in the DFT data of the subsurface layer, but not for the bulk.

Figure \ref{Fig3}a and b show exemplary sketches of the relaxed atomic structure of the bulk unit cell and the unit cell of the slab with Te-terminated surfaces, respectively, according to our DFT calculations. Figure \ref{Fig3}c and d showcase the corresponding radial distribution functions (RDFs) of interatomic distances using all six computational results.
The  peak maxima in bulk RDFs for the different bonding partners (Fig. \ref{Fig3}c) are in good agreement with EXAFS and XRD results \cite{Kolobov2004}. The smallest peak width is observed for the Te-Te distance evidencing the large order in the Te layer. For the strongly discussed Ge-Te bonds \cite{Liu2011,Krbal2012}, the RDF exhibits two peaks, corresponding to the well-known three shorter and three longer Ge-Te bonds \cite{Robertson2007,Shportko2008}. There are no indications of a bond length $d \leq 2.7$ \AA, which would be the hallmark of tetrahedral Ge \cite{Liu2011,Lee2017}.

\begin{figure}
\includegraphics[width=1\linewidth]{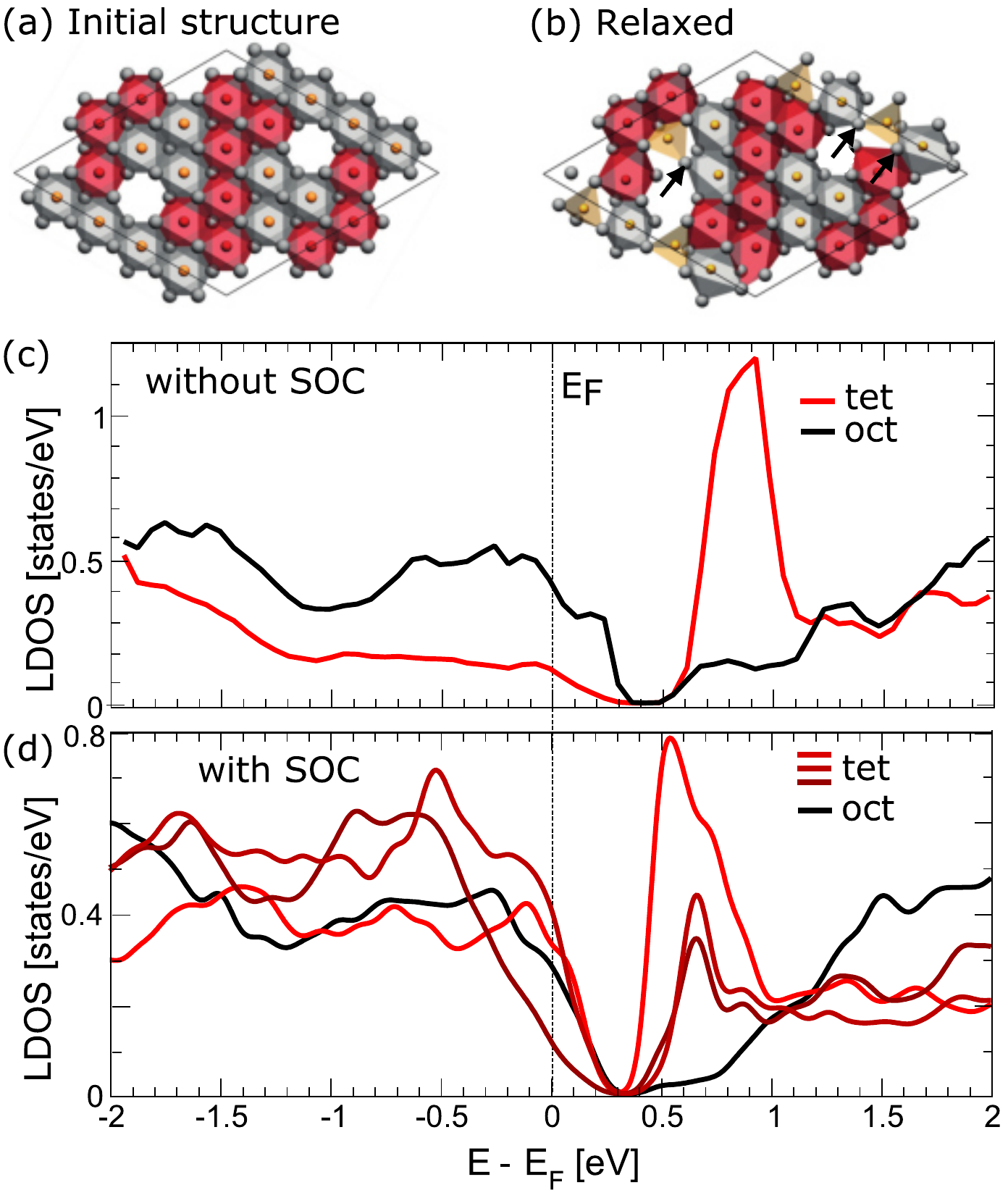}
\caption{(a), (b) Atomic positions of the upper three layers of one of the slabs as in Fig.~\ref{Fig3}b (a) before and (b) after structural relaxation: Te atoms (grey), Sb atoms (red),
  Ge atoms (orange); the coordination of Ge and Sb atoms is highlighted by surrounding, semitransparent polyhedra; octahedral Ge (grey), tetrahedral Ge (orange), octahedral Sb (red); arrows mark atoms used in (d). (c) Calculated LDOS without SOC for s+p+d orbitals integrated over the area of a tetrahedral (red line, tet) and an octahedral (black line, oct) Ge-Te bond. (d) Same as (c), but for the calculation with SOC and after integration over the volume of surface Te atoms, which are next to Ge atoms. The three tetrahedrally bonded Te atoms are marked by arrows in (b). \label{Fig4}}
\end{figure}

 The RDFs of the slabs with Te surfaces (Fig.~\ref{Fig3}d) reveal slightly broader peaks for Te-Te bonds and Sb-Te bonds, highlighting the stronger flexibility of the atoms close to the surface, but without changing peak positions or peak substructures. In contrast, the Ge-Te bond exhibits an additional peak at $d=2.65$ \AA{}  (arrow in Fig. \ref{Fig3}d), which indicates tetrahedrally coordinated Ge \cite{Rosenthal2011}.

 To identify the tetrahedral bonds in real space, coordination polyhedra are added around the subsurface atoms, prior and after relaxation of the slab model (Fig. \ref{Fig4}a, b). The initial structure (Fig. \ref{Fig4}a) consists only of GeTe$_{6}$ and SbTe$_{6}$ octahdera with 6-fold coordination. In contrast, the relaxed structure (Fig. \ref{Fig4}b) also features a number of tetrahedral GeTe$_4$ structures (orange tetrahedra). We checked that these tetrahedra correspond to the bond lengths $d<2.7$ \AA{} (Fig.~\ref{Fig3}d). The subsurface layer, thus, exhibits tetrahedral Ge bonds in contrast to the bulk. This exemplifies that the surface could be different in structure implying differences in other relevant properties. For example, the larger susceptibility of the surface to tetrahedral Ge might be important for the switching propagation at the interface to the amorphous phase, which is known to contain more tetrahedral than octahedral Ge bonds \cite{Sosso2011,Krbal2011,Lee2017}.
 Notice that the tetrahedrally bonded Ge atoms of the metastable film are always adjacent to a vacancy within the Ge/Sb/Vc layer. We assume that this provides the required flexibility of some of the Te bond partners, such that they can move either closer or further apart from the respective Ge atom.

\begin{figure}
\includegraphics[width=1\linewidth]{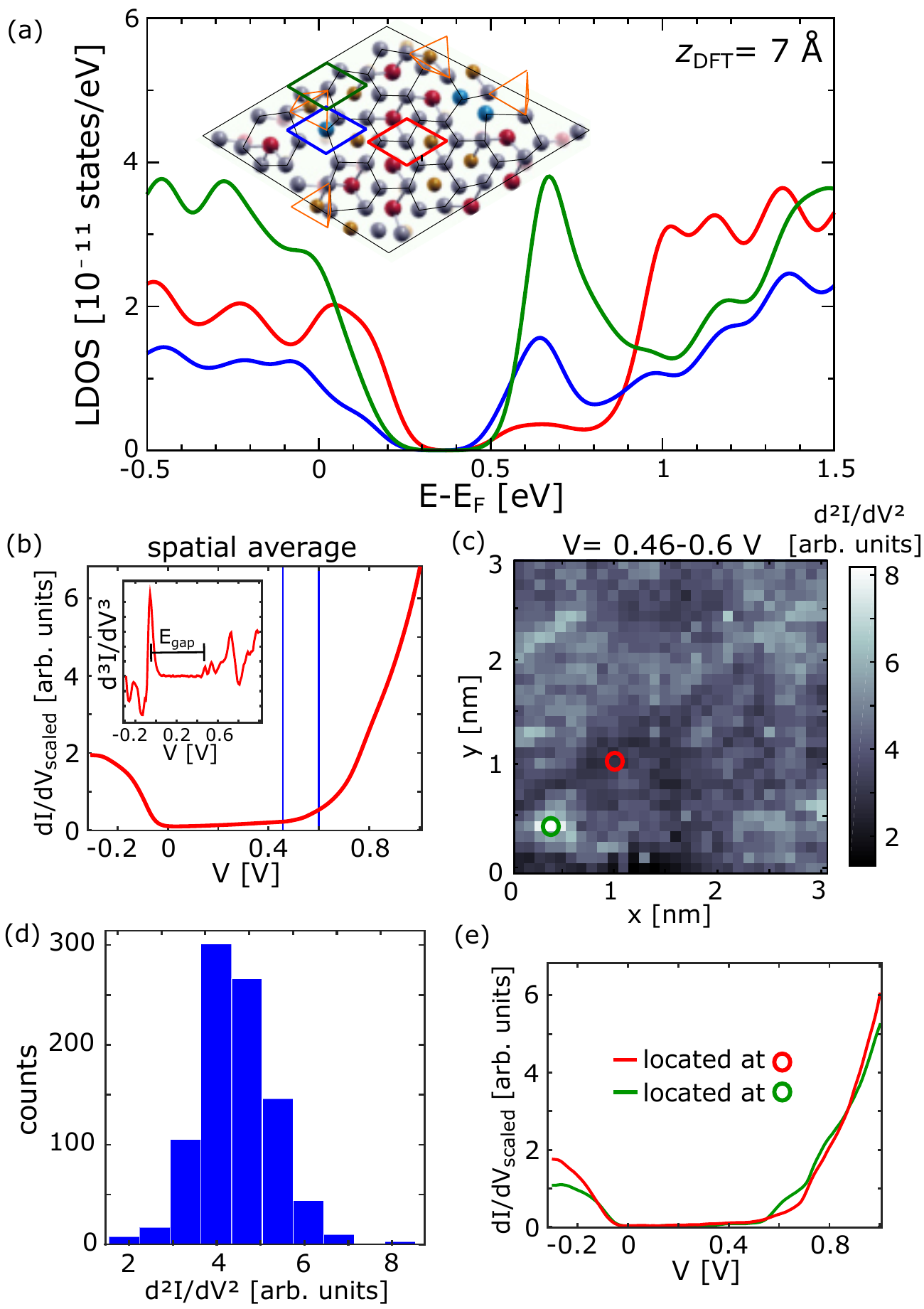}
\caption{(a) Calculated LDOS in vacuum ($z_{\rm DFT}=7$\,\AA) integrated over the equally colored diamonds in the inset, SOC included. Inset: atomic structure of the upper three layers of the relaxed GST-225 slab (same as Fig. \ref{Fig4}b), Te atoms (grey, blue), Sb atoms (red),  Ge atoms (orange), diamonds are centered above octahedrally bonded (red) and tetrahedrally bonded (green, blue) Te; orange pyramids surround tetrahedrally bonded Ge; blue Te atoms are under-coordinated and exhibit a peak at the CBM, too. (b) Spatially averaged $dI/dV(V)$ spectrum, $V_{\rm stab}=-300$\,mV, $I_{\rm stab}=50$\,pA, $T=9$\,K blue lines mark the averaging interval used in (c). Inset: 2$^{\rm nd}$ derivative of the main curve with marked band gap $E_{\rm gap}$. (c) Map of $d^{2}I/dV^{2}(V)$ averaged over $V=0.46-0.6$\,V (blue lines in (b)); minimum and maximum positions are marked by a green and a red circle, respectively. (d) Histogram of $d^{2}I/dV^{2}(V)$ values from (c). (e) $dI/dV(V)$ at the two extremal positions as marked in (c), $V_{\rm stab}=-300$\,mV, $I_{\rm stab}=50$\,pA, $T=9$\,K. \label{Fig5}}
\end{figure}

Fortunately, we find that tetrahedral and octahedral Ge-Te bonds exhibit a distinct LDOS close to $E_{\rm F}$ (Fig. \ref{Fig4}c, d). This implies the possibility to distinguish them by STS.
Figure~\ref{Fig4}c shows two examples of the LDOS integrated over a tetrahedral and an octahedral Ge-Te bond area. A strong peak at the CBM characterizes the tetrahedral bond, which is found similarly for all tetrahedral bonds with peak height variations by up to a factor of three and slightly shifting peak energies by up to $200$\,meV. In contrast, the octahedral bond exhibits a comparatively featureless LDOS at the CBM. Including SOC and integrating the LDOS across the surface Te atom next to the tetrahedral bond changes details of the peak, but not its general appearance (Fig. \ref{Fig4}d). Also within the vacuum area above the partially, tetrahedrally bonded surface Te atoms, the appearance of the peak does not change, at least, up to $z_{\rm DFT}=7$\,\AA{} (Fig. \ref{Fig5}a). The peak mostly consists of Ge 4s- and the Te 5p-orbitals. We find that also under-coordinated Te atoms close to vacancies exhibit a peak at the CBM, even if not involved in a tetrahedral bond (blue atoms in inset of Fig.~\ref{Fig5}a). Hence, the peak is a robust feature of the tetrahedral bond or of vacancies, which should be visible in the STS data.

Surprisingly, however, we never observe such a peak in the experimental $dI/dV$ curves. We also searched for smaller features close to the CBM. Figure~\ref{Fig5}b$-$e exemplifies the searching strategy. Firstly, the spatially averaged $dI/dV$ curve (Fig. \ref{Fig5}b) corroborates the band gap $E_{\rm gap}\simeq 0.5$\,eV  \cite{Lee2005,Pauly2013,Kellner2017} by the peak distance in spatially averaged $d^3I/dV^3$ curves (inset). A distance of the peaks surrounding the gap of $E_{\rm gap}=0.51\pm 0.04$\,eV is found. Secondly, the energy region above the resulting CBM ($V=0.46-0.6$\,eV) is studied in detail, e.g., by determining maps of the slope of $dI/dV$ within this voltage range (Fig.~\ref{Fig5}c). Fluctuations by up to a factor of three are visible (see also histogram in Fig.~\ref{Fig5}d), but the resulting differences of the respective $dI/dV$ curves are small (Fig.~\ref{Fig5}e). These differences resemble the differences that we found above different octahedrally bound Te atoms within the DFT data (not shown).
Hence, we exclude a significant amount of tetrahedral Ge in the subsurface layer.

It is likely that the suppression of tetrahedral bonds is related to the strongly reduced vacancy density in the subsurface layer (Fig. \ref{Fig1}). This prohibits the required relaxation of neighboring Te atoms in order to realize the four shorter bonds of a tetrahedral Ge. In that respect, it would be interesting to investigate sputtered films of GST-225 after UHV transfer, which might exhibit the fingerprints of tetrahedral subsurface Ge in STS due to their reduced tendency for vacancy ordering \cite{Bragaglia2016}.

\section{Individual defect configurations}

\begin{figure}
\includegraphics[width=1\linewidth]{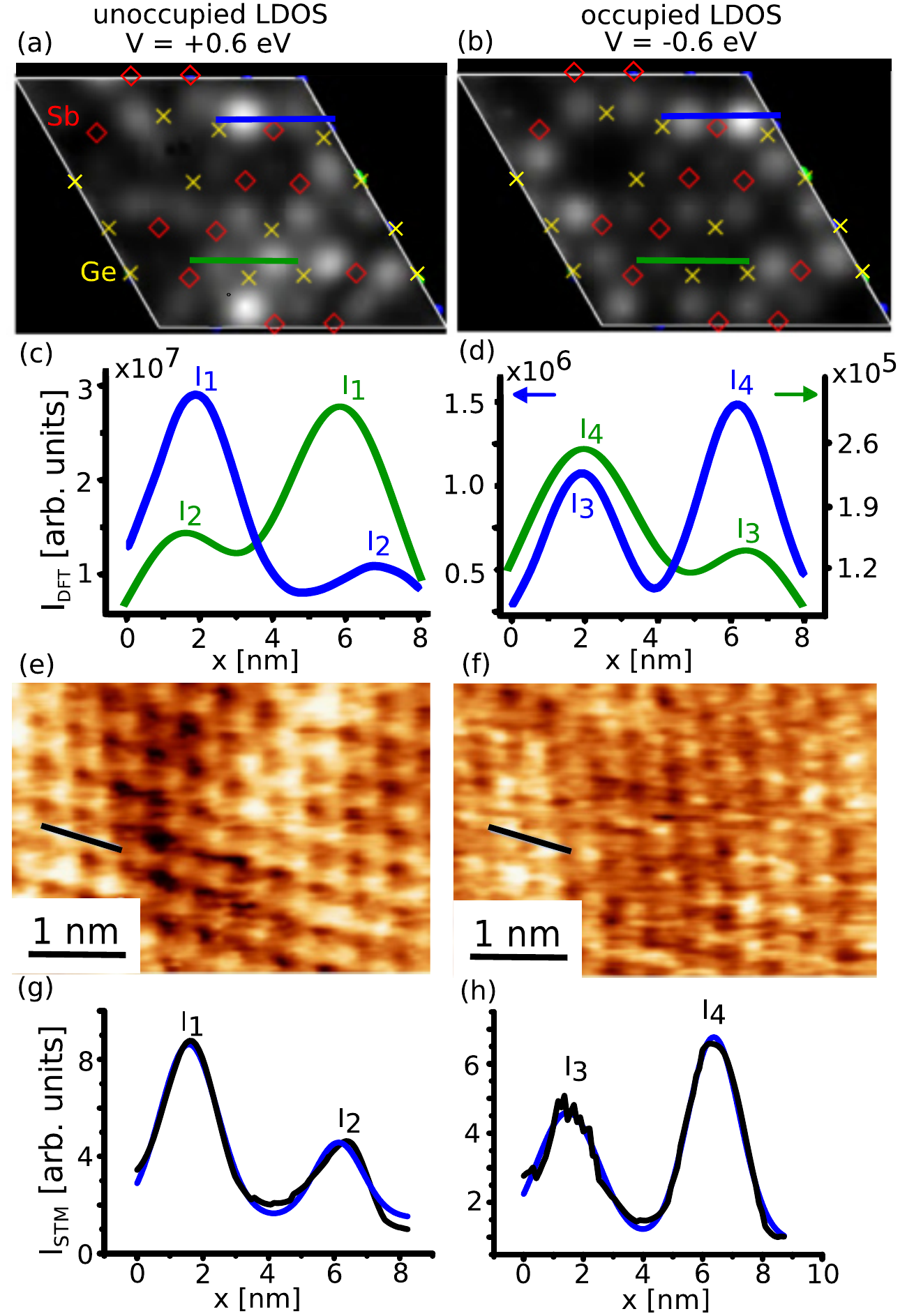}
\caption{(a), (b) Simulated STM images of a Te terminated GST-225 slab with disordered subsurface layer, $z_{\rm DFT}=4$\,\AA, different $V$ as marked, without SOC; positions of Ge atoms (yellow crosses) and Sb atoms (red diamonds) are overlaid; lines mark profile lines shown in (c), (d). (c), (d) Profile lines as marked in (a), (b), respectively; note the three different scales, which are directly comparable; peaks are labeled for comparison with the experimental data in (e)$-$(h). (e), (f) Atomically resolved STM images of the identical area, $ I = 50$\,pA, $T=300$\,K (e) $V = +0.6$\,V,  (f) $V = -0.6$\,V; black lines mark the position of the scaled profile lines in (g), (h). (g), (h) Profile lines along the lines in (e), (f), respectively, after scaling into $I_{\rm STM}$ according to eq. (\ref{eq:2}); Gaussian fit curves (blue) with maxima marked $I_1-I_4$ are added. \label{Fig6}}
 \end{figure}

\begin{figure}
\includegraphics[width=1\linewidth]{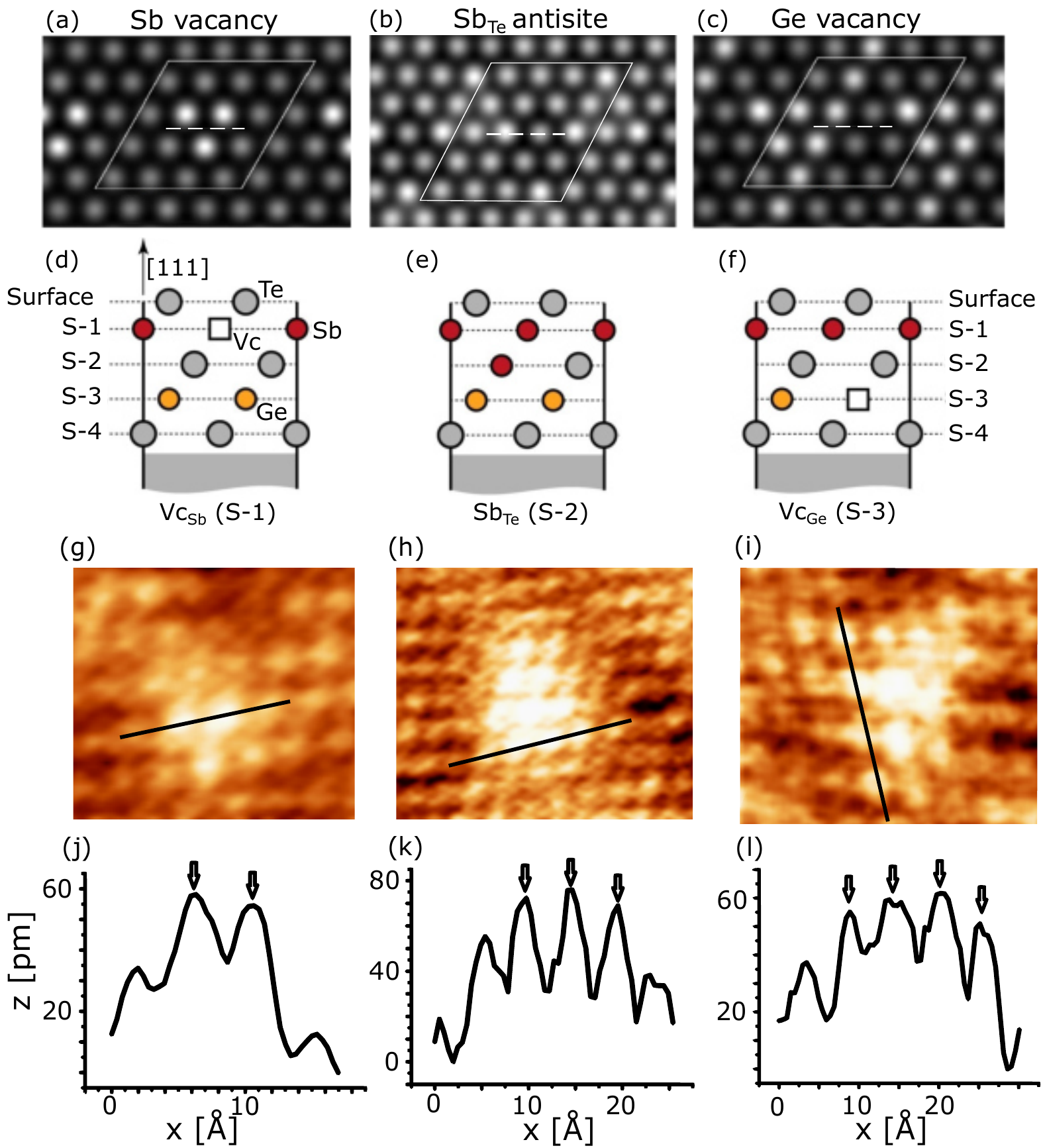}
\caption{(a)$-$(c) Simulated $I_{\rm DFT}(x,y)$ images in logarithmic scale at $z_{\rm DFT} = 4$ \AA{}, $V=-1.0$\,V , for three different defect configurations as sketched in (d)$-$(f), respectively, without SOC; the white parallelogram marks the in-plane unit cell of the calculation (side length $1.68$\,nm); the defects are located in the center of the in-plane unit cell;
dashed lines show the position of cuts in (d)$-$(f). (d)$-$(f) Atomic structure of the defect configurations of a vertical cut along the dashed line marked in (a)$-$(c); atomic symbols are labeled in (d); note that some of the displayed atoms are not centered at the dashed cutting line of (a)$-$(c). (g)$-$(h) STM images of characteristic triangular protrusions, $V=-0.5$\,V, $I=100$\,pA, $T=300$\,K; black lines mark profile lines shown in (j)$-$(l). (j)$-$(l) Profile lines across the protrusions as  marked in (g)$-$(i), respectively; arrows mark the largest maxima along the line. \label{Fig7}}
 \end{figure}

The investigations described above showcase that the subsurface layer of epitaxial, metastable GST-225 is significantly ordered. Hence, we tried to identify individual defects by comparison of the STM data with DFT calculations of an ordered, cubic GST-225 slab that includes only isolated defects (Fig.~\ref{Fig7}d$-$f).

We start with the attempt to single out features in a completely disordered subsurface layer. Figure~\ref{Fig6}a and b show two simulated STM images, i.e., $I_{\rm DFT}:=\int_{E_{\rm F}}^{eV} {\rm LDOS}(E) dE$ in logarithmic scale, at $z_{\rm DFT}=4$\,\AA, for positive and negative $V$, respectively. The positions of subsurface atoms are superimposed. Obviously, the appearance of the STM image depends on the voltage polarity in line with the experimental data (Fig. \ref{Fig6}e$-$f). For example, the intensity relation between two neighboring apparent Te atoms inverts with voltage polarity at the positions below the blue and green line in Fig.~\ref{Fig6}a$-$b. This peak height inversion is more clearly apparent within the profile lines of Fig.~\ref{Fig6}c$-$d, where the maxima are labeled by $I_1-I_4$. A peak height inversion between neighboring Te atoms is also observed in the STM data as shown exemplary in Fig.~\ref{Fig6}e$-$h. The peak intensity ratios in the experiment are deduced by Gaussian fits (blue lines in Fig.~\ref{Fig6}g$-$h) as $I_{1}/I_{2}= 1.9$ at positive $V$ and $I_{3}/I_{4}=0.7$ at negative $V$.
The simulations (Fig.\ref{Fig6}c$-$d) exhibit $I_{1}/I_{2}= 2.9 $~$(2.0)$ and  $I_{3}/I_{4} = 0.8$~$(0.5)$ for the blue (green) profile lines. This constitutes a reasonable agreement.
Experimentally, we find a density of clearly inverting atom pairs, i.e., $I_1/I_2 \ge 1.2$ and $I_3/I_4 \le 0.9$, $n_{\rm inv}= (0.3 \pm 0.1)$/nm$^2$, which is smaller than in the DFT data ($n_{\rm inv}= (0.7 \pm 0.1)$/nm$^2$), but well within the same order of magnitude.

One might be tempted to relate the inverting features to a particular atomic subsurface configuration. However, from the DFT data (Fig.~\ref{Fig6}a$-$b), we deduce directly that the atomic subsurface configuration at the green line is different from the one at the blue line.
Hence, it is impossible to relate the qualitative LDOS feature of inverted relative intensity of two neighboring surface Te atoms to an atomic subsurface configuration. This assignment problem is strongly related to the shear amount of possible bond partner configurations of the two neighboring surface Te atoms, which amounts to $3^5=243$, assuming a complete randomness of Ge, Sb and Vcs in the subsurface layer.

Nevertheless, we find some more isolated corrugation features, which look strikingly similar to energetically favorable defect configurations within an ordered GST-225 slab.
The first type is a Sb vacancy (Vc$_{\mathrm{Sb}}$) in the subsurface layer S-1 ( Fig. \ref{Fig7}a,d), which is surrounded by octahedrally bonded Sb and Te atoms. It has a formation energy of $0.42$ eV. The second type is an antisite defect (Sb$_{\mathrm{Te}}$), where the Sb atom occupies a place in the Te layer S-2, which is two layers below the surface  (Fig. \ref{Fig7}b). This defect exhibits a formation energy of $0.51$ eV. The third type is a vacancy at a Ge place (Vc$_{\mathrm{Ge}}$) in the layer S-3, three layers below the surface, (Fig.  \ref{Fig7}c) with negative formation energy of $-0.32$ eV \cite{Wuttig2006}. We cannot exclude a remaining small interaction between defects in neighboring unit cells due to the limited cell size, which, however, will not change these energies significantly.

All three types of defects lead to triangular features in the corresponding, simulated STM images. The feature size increases with the depth of the defect below the surface as expected. The subsurface Vc leads to three slightly brighter Te atoms  on top (Fig.~\ref{Fig7}a) in line with the observation for subsurface Vcs surrounded by octahedrally bonded Te in Fig. \ref{Fig1}c,e (red triangles). Such structures are also found in the experimental data (Fig.~\ref{Fig7}g,j) indicating that the subsurface layer is not completely depleted of Vcs. Other types of triangular structures are also found in the experiment with side length of $3-4$ Te atoms (Fig.~\ref{Fig7}h,i,k,l) implying that the corresponding defects are located deeper than subsurface.
These triangularly appearing defects are distributed rather homogeneously across the surface (see e.g. Fig. \ref{Fig1} (b)) with an  areal density of $n_{\rm defect}\simeq (1.5\pm 0.5)\cdot 10^{-4}$/nm$^2$. This number could be regarded as an upper bound for the number of subsurface vacancies, which would be $\sim 10^{-4}$ of the subsurface atoms. Note that $10^{-4}$ is exactly the vacancy (acceptor) percentage required to explain the measured charge carrier density $p=3\cdot10^{26}$/m$^3$ of identically prepared samples \cite{Kellner2017}.
More importantly, the rather separated and, thus, dilute defect structures, some of them probably even belonging to defects below the subsurface layer, corroborate that the subsurface area is significantly more ordered than usually expected for the meta-stable rock-salt structure (compare Fig.~\ref{Fig1}c$-$f).

\section{Summary}
Using combined STM and DFT studies, we have explored the subsurface Ge/Sb/Vc layer of epitaxial, metastable GST-225, which appears to become a model system for a more detailed atomistic investigation of the technologically important phase change materials. We confirmed experimentally that the epitaxial films are Te-terminated. Additionally, we reveal that the subsurface layer is much more ordered than expected. In particular, it is strongly depleted of vacancies, which would mostly appear at depressions in the STM images of the Te layer. The alternative appearance of vacancies as a protrusion, in case of being surrounded by octahedrally bonded atoms, has been found with a density of less than $2\cdot 10^{-4}$/nm$^2$, i.e., by three orders of magnitude smaller than expected for a completely disordered subsurface layer. Moreover, we find that the potential fluctuations within the surface layer ($\sim 20$\,meV) are much less than expected from a disordered subsurface layer and are likely explainable by the charged acceptors (Ge vacancies) within the bulk of the system only. Finally, we have shown that a disordered subsurface layer would be prone to tetrahedral Ge bonds, exhibiting a strong peak in the LDOS at the conduction band minimum, which is, however, absent in the epitaxial films due to the increased order. We assume that the absence of vacancies does not allow the required relaxation around the shorter tetrahedral bonds.

\section{Acknowledgment}
We appreciate financial support by the German Science Foundation (DFG): SFB 917, Project A1 and A3; V.L.D. thanks the German Academic Scholarship Foundation for a doctoral fellowship; G.B. gratefully acknowledges the computing time provided on the JARA-HPC Partition part of the supercomputer JURECA at Forschungszentrum J\"ulich. The work was partly supported by the Leibniz Gemeinschaft within the Leibniz Competition on the project {\it Epitaxial phase change superlattices designed for investigation of non-thermal switching}.

%

\end{document}